\documentclass[aps,prb,twocolumn,groupedaddress,showpacs]{revtex4}
\usepackage{epsfig}
\usepackage{graphicx}
\begin{document}

\title{Determinant Quantum Monte Carlo Study of the
Screening of the One Body Potential
near a Metal-Insulator Transition}
\author{P. B. Chakraborty$^1$, P. J. H. Denteneer$^2$, 
and R. T. Scalettar$^1$}
\affiliation{$^1$Physics Department, University of California, 
Davis, California 95616, USA}
\affiliation{$^2$ Lorentz Institute, LION, Leiden University,
P.O. Box 9506, 2300 RA Leiden, The Netherlands}

\begin{abstract}
In this paper we present a determinant quantum monte carlo study 
of the two dimensional Hubbard model with random site disorder.
We show that, as in the case of bond disorder, the system undergoes
a transition from an Anderson insulating phase to a metallic phase as the
on-site repulsion $U$ is increased beyond a critical value $U_c$.
However, there appears to be no
sharp signal of this metal-insulator transition in the screened
site energies.  We observe that, while the system remains metallic
for interaction values up to twice $U_c$, the conductivity is maximal
in the metallic phase just beyond $U_c$, and decreases for larger
correlation.
\end{abstract}

\pacs{
71.10.Fd, 
71.30.+h, 
02.70.Uu  
}
\maketitle

\section*{Introduction}

The metal-insulator transition arising from the competition of
randomness and interactions remains an intriguing problem in
condensed matter physics.  For example, the question of the existence of a
metallic phase in two dimensions, for which an experimental consensus
had emerged in the 1980's \cite{old2dmitrev}
has been revisited with new samples over the last decade,
with developments which have driven
a considerable amount of new theoretical work \cite{new2dmitrev,punnoose05}.  

Several interesting lines of study have emerged which explore the interplay
of one-body potentials and two-body interactions in more general contexts.
The superfluid-insulator transition has been studied in disordered,
interacting boson systems, where the existence of a thermodynamic 
order parameter, the superfluid density $\rho_s$, as well as the greater
ease of numerical simulations, has resulted in many definitive 
results \cite{fisher89,scalettar91,krauth91}. The coexistence of
a metal and a Mott-Hubbard insulating phase in the disordered
half-filled Hubbard model has been explored using both numerical
and analytical techniques \cite{aguiar05}.
The existence of insulating phases away from commensurate fermion
filling has been explored in models with bimodal distributions
of on-site chemical potential \cite{byczuk03,byczuk04}.
Finally, the question of metallic phases
arising from the addition of correlations to a {\it band} insulator
is drawing new attention \cite{garg06}.

The commonly cited qualitative picture of the appearence of
a metallic phase out of a disordered one is that the interactions
act to screen the one-body potential.
While several Quantum Monte Carlo studies of disordered interacting
fermions exist \cite{denteneer99,denteneer01,lee04} which
demonstrate the possibility of a metallic phase, none have
looked quantitatively at this screening in the Hubbard hamiltonian.

In this paper, we will present results for the conductivity and
renormalized site energy of the two dimensional Anderson-Hubbard model,
\begin{eqnarray}
H=&-&t \sum_{\langle jl \rangle,\sigma}
(c^{\dagger}_{j\sigma}c_{l\sigma}+ c^{\dagger}_{l\sigma}c_{j\sigma})
\nonumber
\\
&+&U \sum_j n_{j\uparrow}n_{j\downarrow} +
\sum_j (\epsilon_j - \mu) (n_{j\uparrow}+n_{j\downarrow})
\label{AndersonHubbard}
\end{eqnarray}
Here $c^{\dagger}_{j\sigma},(c_{j\sigma})$ are 
fermion creation(destruction) operators
on site $j$ for spin $\sigma$ and 
$n_{j\sigma}=c^{\dagger}_{j\sigma}c_{j\sigma}$ is the number operator.  
$t$ is the hopping parameter, $U$ the onsite repulsion,
and $\mu$ and $\epsilon_j$ the global chemical potential and local
site energies respectively.  
Each $\epsilon_j$ is drawn independently from a uniform distribution
on $(-\frac12 \Delta,+\frac12 \Delta)$.
We choose $t=1$ to set our scale of energy.

Our key conclusion is that while increasing $U$ can drive an 
Anderson insulating phase metallic, there appears to be no sharp
signature of this transition in the variance of the
renormalized site energies.  This suggests that the metallic phase
arises at least partially from an additional mechanism beyond
a simple screening of the one-body potential.

\section*{Numerical Approach}

We employ the determinant quantum monte carlo (DQMC) method \cite{blankenbecler81}.
Since many descriptions of the approach exist, we only provide
a brief sketch here, focusing on those features most relevant to
the present study.  DQMC is an exact method to compute
the properties of tight binding Hamiltonians on finite lattices.
The inverse temperature $\beta$ in the partition function is discretized and
an auxiliary (``Hubbard-Stratonovich") field introduced to decouple
the interactions.  The resulting quadratic form in fermion creation
and destruction operators is integrated out analytically, leaving
a sum over the Hubbard-Stratonovich variables which can be performed
stochastically.

We have chosen the imaginary time discretization size
small enough such that the systematic ``Trotter'' errors are comparable
to the statistical errors associated with the monte carlo sampling and disorder averaging.
Of greater concern in these simulations is the finite lattice size
and, in particular, the possibility of a ``false" signal of metallic
behavior which would occur if the localization length exceeds the
lattice size.  We have verified that in the phases we identify as
metallic the localization length (computed at $U=0$) 
is less than the lattice size.

To investigate the metal-insulator transition, we look directly at
the dc conductivity which we obtain from the current-current correlation function
\begin{equation}
j_x ({\bf \ell},\tau)=  e^{H \tau} [\,i t \sum_\sigma \, 
(c^{\dagger}_{{\bf \ell} + \hat{x},\sigma}
c^{\phantom \dagger}_{{\bf \ell}\sigma}
- c^{\dagger}_{{\bf \ell}\sigma}
c^{\phantom \dagger}_{{\bf \ell}+\hat{x},\sigma}) \,]\ e^{-H \tau} .
\label{cccf}
\end{equation}
We compute the Fourier transform 
$j_x ({\bf q},\tau)$ 
of
$j_x ({\bf \ell},\tau)$  and its correlation function
$\Lambda_{xx} ({\bf q},\tau) = \langle j_x ({\bf
q},\tau) \, j_x (-{\bf q}, 0) \rangle$.
Using the general formalism of linear response theory, the dc conductivity is
given by
\begin{equation}
\sigma_{\rm{dc}}={\rm{lim}}_{\omega\rightarrow 0}\frac{{\rm{Im}}\Lambda_{xx}({\bf q}=0,\omega)}{\omega}
\label{linearresponse}
\end{equation} 
The frequency dependent conductivity is given by the integral transform
\begin{equation}
\Lambda_{xx} ({\bf q},\tau)
=\int_{-\infty}^{\infty}\frac{d\omega}{\pi}\frac{e^{-\omega \tau}}{1-e^{-\beta \omega}}
{\rm{Im}}\Lambda_{xx} ({\bf q},\omega)
\label{kernel}
\end{equation}

It is difficult to obtain ${\rm{Im}}\Lambda_{xx}({\bf q},\omega)$ by inverting this integral equation,
because it requires the determination of $\Lambda_{xx}({\bf q},\tau)$ on a very fine mesh of imaginary times $\tau$
with very high numerical accuracy. However, if we insert $\tau=\beta/2$ in Eqn.~\ref{kernel}, the function multiplying
${\rm{Im}}\Lambda_{xx}({\bf q},\omega)$ for low $T$ effectively restricts the integral to small $\omega$, so that we may approximate
${\rm{Im}}\Lambda_{xx}({\bf q}=0,\omega)$ by $\sigma_{\rm{dc}}\omega$. The frequency integral may now be evaluated analytically, leading to the result:
\begin{equation}
 \sigma_{\rm dc} = 
   \frac{\beta^2}{\pi} \Lambda_{xx} ({\bf q}=0,\tau=\beta/2) 
 \label{eq:condform}
\end{equation}

This approximation is expected to be valid when the temperature is smaller than an appropriate energy scale in the problem.
It is convenient because it allows the computation of $\sigma_{\rm{dc}}$ as a function of temperature to be obtained from the function
$\Lambda_{xx}({\bf q},\tau)$, which is calculated directly in DQMC.

Obtaining a transport property like $\sigma_{\rm dc}$
directly from imaginary time data, as described above,
is a process which must be undertaken with caution.
However, the use of this procedure gives the correct
characterization as a metal or insulator in all cases which 
we have checked so far.  For example, 
$d\sigma_{\rm dc}/dT$ is positive (insulating behavior)
for the half-filled $d=2$ Hubbard model without randomness at
all values of $U$, that is, regardless of whether the
insulating character arises predominantly from
antiferromagnetic order (weak coupling) or Mott behavior 
(strong coupling) \cite{denteneer99,paris06}.
The procedure also gives the correct physics in a band-insulator
when a staggered site energy is present and $U=0$. It has also been 
shown to give the correct physics of the disordered attractive Hubbard
model\cite{trivedi96}.

A fundamental check of the numerical data
is the verification that the longitudinal
current-current correlation function obeys the gauge invariance condition,
\begin{equation}
\Lambda_{xx} (q_x\rightarrow 0\,,q_y=0\,,i\omega_n=0) = {\cal K}
\label{GIC}
\end{equation}
where ${\cal K}$ is the kinetic energy.
We have checked that,
as in previous work \cite{scalapino93,trivedi96,denteneer99}
this condition is satisfied.


Since the disordered site-energies in the system are generated randomly from 
a uniform distribution $(-\frac{1}{2}\Delta,+\frac{1}{2}\Delta)$, the distribution has
zero mean, and a variance:

\begin{equation}
{\cal{V}}_{0}^{2}=\frac{1}{\Delta}\int_{-\frac{\Delta}{2}}^{\frac{\Delta}{2}}\epsilon^2 d\epsilon=\frac{\Delta^{2}}{12}
\label{v0}
\end{equation}

In order to study the screening of the disordered
potential by interactions we note that within a mean-field picture an
electron moving in the one body potential $\epsilon_j$ will 
feel the site energy renormalized by the density of oppositely
oriented electrons. 
That is,
\begin{equation}
\tilde{\epsilon}_{j,\sigma} =\epsilon_j + U \langle n_{j,-\sigma}\rangle 
\label{rse}
\end{equation}
which becomes, in the absence of spin-polarization,
\begin{equation}
\tilde{\epsilon}_{j}=\epsilon_j + \frac{U}{2}\langle n_{j}\rangle
\label{nospinpol}
\end{equation}
since for each spin species $\langle n_{j\sigma}\rangle = \frac{1}{2} \langle n_{j}\rangle $.
We define an associated dimensionless variance
by normalizing to the fluctuations in the original site energies,
\begin{equation}
{\cal{V}}^2=\frac{1}{{\cal{V}}_0^2}(\langle \tilde{\epsilon}_{j}^{2}\rangle - \langle \tilde{\epsilon}_{j}\rangle^{2}) ={12 \over \Delta^2 }(\langle \tilde{\epsilon}_{j}^{2}\rangle - \langle \tilde{\epsilon}_{j}\rangle^{2}) 
\label{rsev}
\end{equation}
In the absence of interactions ($U=0$), or for very large $\Delta$ at fixed $U$, we have
${\cal{V}}=1$, indicating that there is no screening of the 
random potential.
The question we wish to address is whether there is some signal,
e.g. a noticeable decrease, in ${\cal{V}}$ on entry
into a metallic phase.

\section*{Metallic phase due to interactions}

\begin{figure}[htp]
\resizebox{\hsize}{!}{\includegraphics[clip=true]{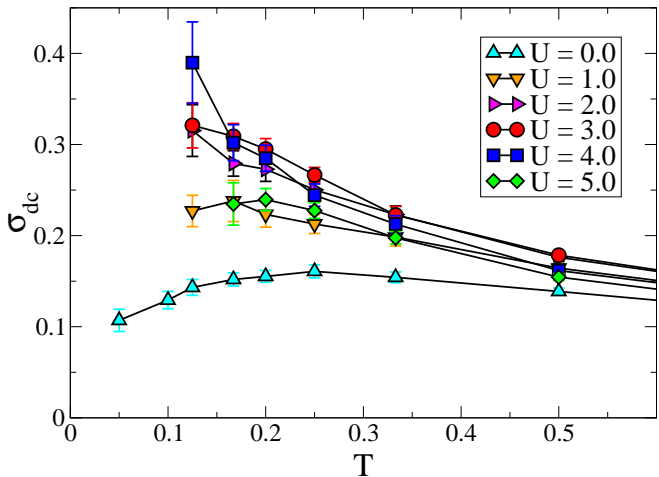}}
\caption{({\bf Color online})The dc conductivity as a function of temperature for increasing values
of the on-site repulsion $U=0-5$.  The site energy variance
$\Delta=8$. } 
\label{fig:Ud8}
\end{figure}

We begin by demonstrating that interactions can drive an Anderson
insulating phase metallic.  We show in Fig.~\ref{fig:Ud8} the dc conductivity as
a function of temperature for a fixed strength of the disorder potential
and increasing $U$.  At $U=0$,
$\sigma_{\rm dc}$
decreases as $T$ is lowered, indicating insulating behavior.  
However, at strong coupling 
$\sigma_{\rm dc}$
increases as $T$ is lowered, indicating a cross-over to
metallic behavior. 
All results in Fig.~\ref{fig:Ud8}, and subsequently in this paper, are at one-quarter
filling $\rho=\langle n \rangle =\frac12$.  This is far away from
the most dramatic effects of $U$ in the Hubbard model-  the Mott
transition and antiferromagnetic ordering.

\begin{figure}[htp]
\resizebox{\hsize}{!}{\includegraphics[clip=true]{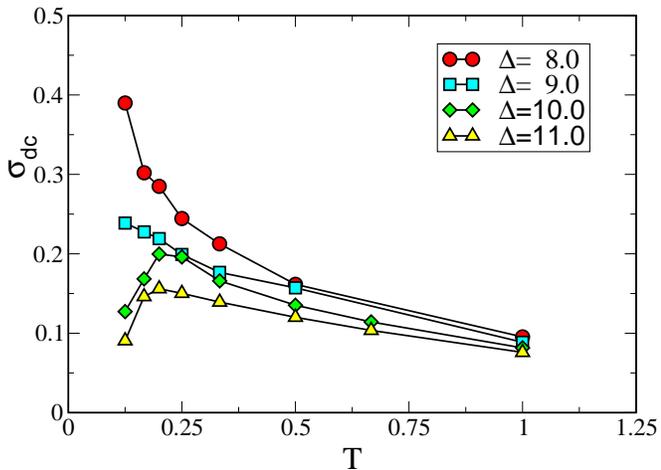}}
\caption{({\bf Color online})The dc conductivity as a function of temperature for increasing values
of disorder $\Delta=8-11$.  The on-site repulsion $U=4$. } 
\label{fig:sigmaall2}
\end{figure}

The metallic phase in Fig.~\ref{fig:Ud8} can be caused to return to
insulating behavior by increasing the site disorder.  This is
shown in Fig.~\ref{fig:sigmaall2} where we begin with the interaction
strength which gives the largest
conductivity, $U=4$, and make $\Delta$ larger.
For $9 < \Delta < 10$ the low temperature slope of
$\sigma_{\rm dc}$
reverts to insulating character.

\begin{figure}[htp]
\resizebox{\hsize}{!}{\includegraphics[clip=true]{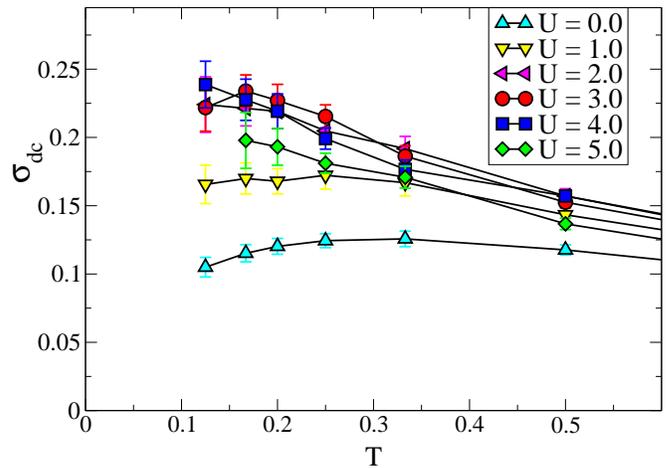}}
\caption{({\bf Color online})Like Fig.~\ref{fig:Ud8}, except at larger disorder, $\Delta=9$.
The same decrease of conductivity with $U$ in the
metallic phase is seen as in Fig.~\ref{fig:Ud8}. } 
\label{fig:Ud9}
\end{figure}

An interesting feature of Fig.~\ref{fig:Ud8} is the non-monotonic behavior of
the conductivity.  
$\sigma_{\rm dc}$
increases with $U$ up to $U \approx 3-4$, but then comes down 
again at $U=5$.  In order to verify that this phenomenon
is generic, we show in Fig.~\ref{fig:Ud9} data for larger
$\Delta=9$.  We again see that 
$\sigma_{\rm dc}$
comes down at strong coupling.
A similar phenomenon occurs in the evolution
of the superfluid density $\rho_s$ for correlated bosons moving in a random
potential- a superfluid phase with $\rho_s \neq 0$ 
exists at intermediate coupling,
but the system is insulating, $\rho_s=0$, both at weak and strong 
coupling \cite{scalettar91}.

\begin{figure}[htp]
\resizebox{\hsize}{!}{\includegraphics[clip=true]{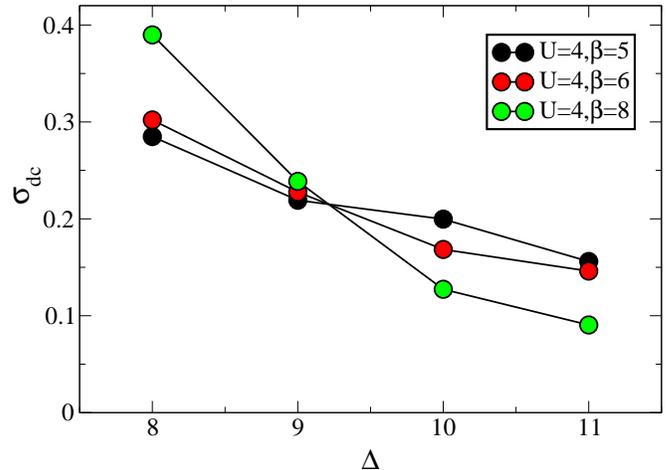}}
\caption{({\bf Color online})A crossing plot for $\sigma_{\rm{dc}}$ vs. $\Delta$. The critical
disorder strength $\Delta_{c}\sim 9.2-9.3$ for $U=4$ is clearly seen. } 
\label{fig:crU4}
\end{figure}

In Fig.~\ref{fig:crU4} we show $\sigma_{\rm{dc}}$ vs. the disorder strength for
progressively lower temperature values. For $\Delta < \Delta_{c}$, the system is metallic and $\sigma_{\rm{dc}}$ increases as the temperature is lowered, while
for $\Delta > \Delta_{c}$, in the insulating state, the behavior is opposite.
The crossing point of the plots demarcates the critical disorder strength.

\begin{figure}[htp]
\resizebox{\hsize}{!}{\includegraphics[clip=true]{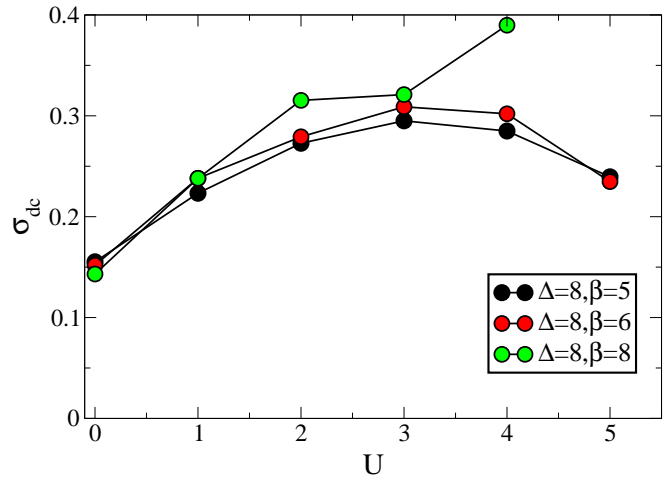}}
\caption{({\bf Color online})A crossing plot for $\sigma_{\rm{dc}}$ vs. $U$. The crossing is
seen to happen for $0 < U <1$ for $\Delta=8$. } 
\label{fig:crd8}
\end{figure}

In Fig.~\ref{fig:crd8}, we show a similar crossing plot for $\sigma_{\rm{dc}}$ as we tune the interaction strength
through the metal-insulator transition for a fixed disorder strength $\Delta=8$ (See Fig.~\ref{fig:Ud8}). A small
value of the interaction is seen to be enough to cause the transition to a metal. Interestingly, the conductivity
is non-monotonic and decreases for large values of the interaction strength (The fermion sign problem in DQMC simulations
forbids the evaluation of $\sigma_{\rm{dc}}$ at $U=5,\beta=8$). It is {\it{possible}} that there is a crossing at larger interaction strengths, when the system reverts back to an insulator. Such non-monotonic behavior behavior of the conductivity has also been seen
in recent DQMC studies of a multi-band Hubbard model at half-filling , where the sequence of transitions with increasing $U$ is found to be Band insulator $\rightarrow$ metal $\rightarrow$ Mott insulator.

      
\section*{Renormalized site-energies with interactions}

\begin{figure}[htp]
\includegraphics[clip=true,height=5.5cm,width=9.3cm]{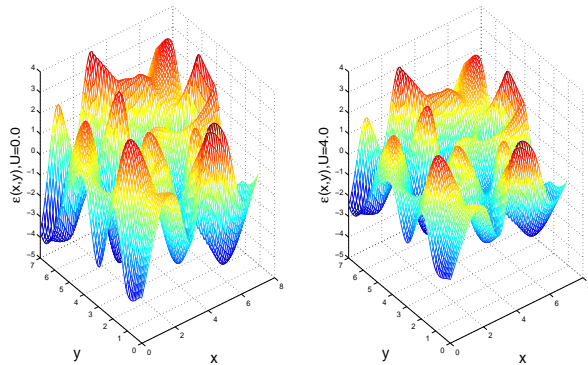}
\caption{({\bf Color online})Left: Landscape of the original site energies $\epsilon_j$ with $\Delta=8$.
Right: Landscape of the renormalized site energies $\tilde{\epsilon}_{j}$ with $\Delta=8$, $U=4$ and
$\beta=8$. On the right, the mean increase in the renormalized site energies due to $U$ has been subtracted
out. } 
\label{fig:demeaned}
\end{figure}

Before showing results for the dimensionless variance of the
renormalized site energies, we present in Fig.~\ref{fig:demeaned} a plot of the
original and renormalized site energy landscapes.  As expected,
the particles preferentially sit on the sites with low $\epsilon_j$,
and these larger values of $\langle n_j \rangle$ then lead to a 
smoother 
$\tilde \epsilon_{j} = \epsilon_j + {U \over 2 } \langle n_{j} \rangle$. However,
there is certainly no very dramatic leveling of the landscape. Below, we will explore this more quantitatively. 

\begin{figure}[htp]
\resizebox{\hsize}{!}{\includegraphics[clip=true]{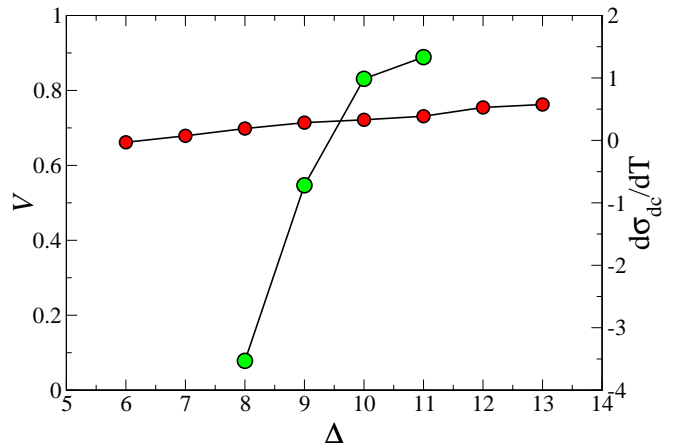}}
\caption{({\bf Color online})The variance of the renormalized site energy ${\cal{V}}$
is shown as a function of $\Delta$, as is the low temperature
slope of the conductivity ($U=4$, $\beta=8$).}
\label{fig:vardsdT}
\end{figure}

In Fig.~\ref{fig:vardsdT} we examine whether there is signal of  the metal insulator
transition in the evolution of ${\cal{V}}$.  We plot the 
low temperature slope $d \sigma_{\rm dc}/dT$ from the data
of Fig.~\ref{fig:sigmaall2} and show its change of sign at $\Delta \approx 9.2-9.3$.
There is no clear indication of this
critical value in the renormalized site energy variance ${\cal{V}}$.

\begin{figure}[htp]
\resizebox{\hsize}{!}{\includegraphics[clip=true]{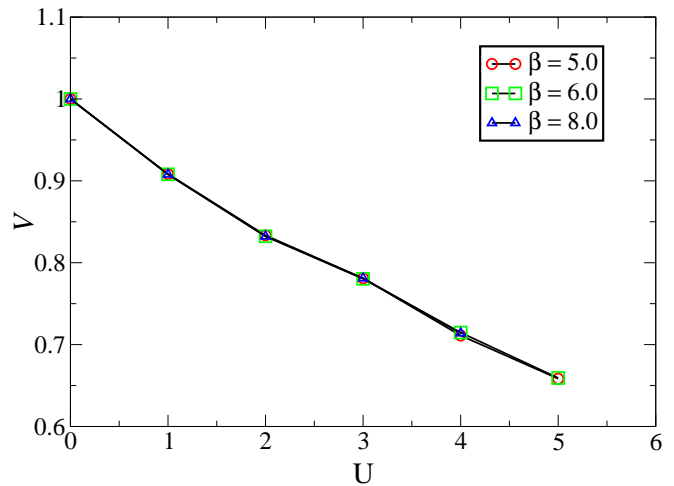}}
\caption{({\bf Color online})The renormalized site energy is shown as a function of $U$.  There
appears to be no signal of the MIT at small $U$, nor the conductivity
peak at $U\approx3-4$.}
\label{fig:Fig6}
\end{figure}

We can similarly look for this effect at the metal-insulator transition
driven by increasing $U$ at fixed $\Delta=8$ (Fig.~\ref{fig:Ud8}).  This is
shown in Fig.~\ref{fig:Fig6}.  Again, there appears to be no clear signal of
the metal-insulator transition in the screened site energies.

\section*{Renormalized site-energies with zero hopping}

The results from the previous section suggest we look more closely at the physical
picture of the smoothening of the site-energy landscape by interactions. Our expectation in
Fig.~\ref{fig:vardsdT} where we plotted ${\cal{V}}$ as a function of the disorder strength 
was that in the metallic phase at weak disorder there would be a markedly smaller value of ${\cal{V}}$,
and then a crossover to a larger value as the disorder is increased into an insulating phase.
On the other hand, at weak disorder we expect the least inhomogeneity in the site occupations.
In the limit of uniform density, at very weak disorder, the site energy variance equals the original one,
and ${\cal{V}}^2=1$, suggesting that ${\cal{V}}$ might instead ${\it{decrease}}$ with disorder. The conflicting
tendency to decrease with disorder as charge-inhomogeneity develops, and increase in the insulating phase, might 
explain why the site energy variance is so insensitive to site energy disorder, whereas when we tune through the transition with interaction strength there is a much larger decline (though still no abrupt signal at the transition). In this section, we examine an analytically solvable limit of the disordered Anderson-Hubbard model, that of $t=0$, which can be considered to be the limit of very high disorder.
At this limit, there is no metallic behavior, but it is still interesting to investigate the behavior of the site-energy distribution as we move from weak to strong interaction (or, strong to weak disorder).

\begin{figure}[htp]
\includegraphics[clip=true,width=4.5cm,height=5.0cm]{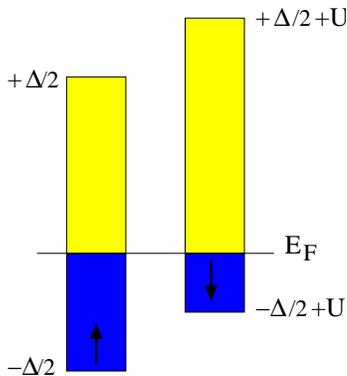}
\caption{({\bf Color online})At $t=0$ the energy levels for occupation by the {\it{first}} electrons (which we denote by
$\uparrow$) extend from $-\Delta/2$ to $\Delta/2$ (left). The energy levels for occupation by the {\it{second}} electrons (which
we denote by $\downarrow$) extends from $-\Delta/2 + U$ to $+\Delta/2 + U$ (right).}
\label{fig:renormE}
\end{figure} 

When there is no hopping, the Anderson-Hubbard model is classical. 
As electrons are added to the lattice at $t=0$, the sites with the
lowest site energies are singly occupied up the Fermi
energy $E_F$.  When, however,  $E_F$ exceeds $-\Delta/2+U$ it becomes
preferable to start doubly occupying the low energy sites.
This is illustrated in Fig.~\ref{fig:renormE}.
From the figure it is evident that 
$\langle n_\uparrow \rangle = {1 \over \Delta} (E_F+\Delta/2)$ and
$\langle n_\downarrow \rangle = {1 \over \Delta}
(E_F+\Delta/2-U)$, and hence that
$\langle n \rangle = \langle n_\uparrow + n_\downarrow  \rangle
= {1 \over \Delta} (2 E_F+\Delta-U)$.
We can easily obtain the mean of the renormalized site energies by
averaging Eqn.~\ref{rse}:
$\langle \tilde \epsilon_\uparrow \rangle
+ \langle \tilde \epsilon_\downarrow \rangle= U (\langle n_\uparrow \rangle + \langle n_\downarrow  \rangle)
= {U \over \Delta} (2 E_F+\Delta-U)$.

A completely equivalent result is obtained by recognizing that
the energies of the sites on which up spin electrons 
reside are raised by $U$
in the range from $-\Delta/2$ to $E_F-U$, where down spins are present.
Similarly,
the energies of the sites on which down spin electrons
reside are raised by $U$
in the range from $-\Delta/2$ to $E_F$, where up spins are present.
(Here the designations `up' and `down' merely reflect the `first' and
`second' electrons on a site.)
When the energies are averaged over these ranges, the same result for
$\langle \tilde \epsilon_\uparrow \rangle
+ \langle \tilde \epsilon_\downarrow \rangle $ is obtained.

The average of the square of the renormalized site energies
is obtained in the same way.  We can then evaluate the dimensionless variance
of the renormalized site energies ${\cal V}_{t=0}^2$, defined as
\begin{equation}
{\cal V}_{t=0}^{2}=\frac{12}{\Delta^{2}}(\langle{\tilde{\epsilon_{j}}^2}\rangle -\langle{\tilde{\epsilon_{j}}}\rangle^{2})
\label{t0}
\end{equation}
To determine the variance, we must distinguish between two cases: a {\it{generic}} one, in which the Fermi
energy $E_{F}$ is larger than $-\Delta/2 + U$, and there is double occupancy of the low-energy sites, and a {\it{non-generic}}
case, in which there is only single occupancy, which may happen for a large value of $x=U/\Delta$ or a small density.
For the generic case, the variance can be computed in terms of the three energy-scales ($U$,$\Delta$ and $E_F$) in the $t=0$ problem:
\begin{eqnarray}
{\cal V}_{t=0}^{2}& = & 1 + \frac{3U^2}{\Delta^2} - \frac{3U}{\Delta}\nonumber\\
& &-\frac{3U^4}{\Delta^4} + \frac{6U^3}{\Delta^3} + \frac{12E_{F}^2U}{\Delta^3}\nonumber\\
& &+\frac{12E_{F}U^3}{\Delta^4}-\frac{12E_{F}U^2}{\Delta^3}-\frac{12E_{F}^2U^2}{\Delta^4}
\label{genvart0}
\end{eqnarray}

The Fermi energy (for the generic case) can be determined in terms of $U$ and $\Delta$
by the following equation:
\begin{equation}
\rho = \frac{E_{F}+\frac{\Delta}{2}}{\Delta} + \frac{E_{F}+\frac{\Delta}{2}-U}{\Delta}
\label{howtogetEF}
\end{equation}
where $\rho$ is the filling. For example, in the quarter-filled case, $\rho=\frac{1}{2}$,
and $E_{F}=\frac{1}{2}(U-\frac{\Delta}{2})$. In the non-generic case, the Fermi energy can,
of course, be determined by simple state counting. 

\begin{figure}[htp]
\resizebox{\hsize}{!}{\includegraphics[clip=true]{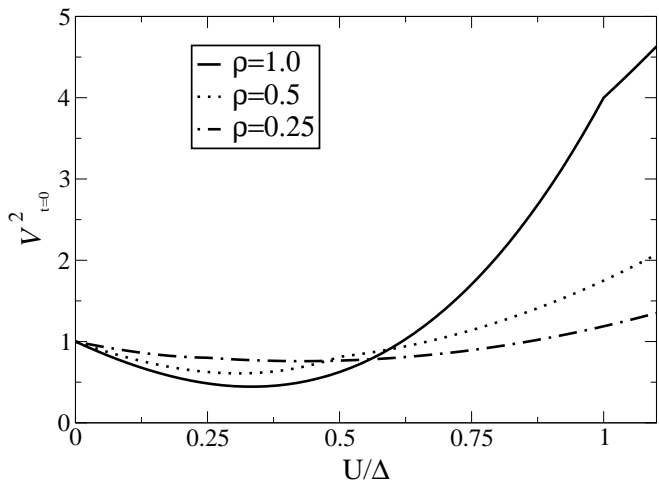}}
\caption{The site-energy variance, as defined in Eq.~\ref{t0}, is plotted for
three different values of the filling, $\rho=1$, $0.5$ and $0.25$. The x-axis is the ratio of the interaction to the disorder strength, $U/\Delta$.}
\label{fig:t0}
\end{figure}

A plot of ${\cal V}_{t=0}^{2}$ vs. $x=U/\Delta$ is given in Fig.~\ref{fig:t0} for different fillings.  
At very weak interaction (or, very strong disorder), the variance equals the non-interacting value 1. 
As the interaction is increased (equivalently, the disorder is decreased), the variance first decreases
and reaches a minimum. Upon further increasing the interaction, however, the site occupations become homogeneous and the variance
grows.

\section*{Conclusions}

We have examined the metal-insulator transition in the
Anderson-Hubbard model using determinant Quantum Monte Carlo
simulations.  Our focus has been on the evolution of the
renormalized site energy through the transition, and
we conclude that it exhibits no sharp feature there. 
It seems that the picture of screening of the disorder
by interaction is too primitive to account for metallic behavior.

On the other hand, we observe an interesting non-monotonic behavior of
the conductivity with interaction strength.  
In the boson Anderson-Hubbard model, the ground state at incommensurate
densities is an Anderson insulator at weak $U$ and a insulating
``Bose glass'' at large $U$.  In between, there is a superfluid
phase in which the superfluid density
first rises as one emerges from the Anderson insulator and then falls
to zero again upon entry into the Bose glass.
As far as we can see, the fermion Hubbard model 
remains metallic at large $U$, but simulations there are difficult
and we cannot make a definitive statement.  In any case, the non-monotonic
behavior of $\sigma_{\rm dc}$ is rather analogous to the behavior
seen for strongly interacting, disordered Bose systems.

We acknowledge support from the National Science Foundation under awards
NSF DMR 0312261 and NSF DMR 0421810. We are grateful to W. E. Pickett, B. Altman and
V. Dobrosavljevic for useful discussions.

\end{document}